\documentclass{ws-ijmpe}
\usepackage{amsmath,amssymb}

\newcommand{\Slash}[1]{\ooalign{\hfil/\hfil\crcr$#1$}}
\newcommand{\bgfig}{\begin{figure}[th]\begin{center}}
\newcommand{\enfig}{\end{center}\end{figure}}
\newcommand{\im}{\mathrm{Im}~}
\newcommand{\re}{\mathrm{Re}~}
\newcommand{\reta}{^\mathrm{R}}

\newcommand{\tr}{\mathrm{Tr}}

\begin{document}
\title{FURTHER ANALYSIS OF EXCITATIONS OF QUARKS AT FINITE TEMPERATURE\\
-- MASS EFFECT AND POLE STRUCTURE --}

\markboth{K.~Mitsutani et al.}
{Further Analysis of Collective Excitations of Quarks at Finite Temperature}

%%%%%%%%%%%%
\catchline{}{}{}{}{}
%%%%%%%%%%%%%%

\author{\footnotesize KAZUYA MITSUTANI\footnote{E-mail:
kazuya@yukawa.kyoto-u.ac.jp}}

\address{Yukawa Institute for Theoretical Physics,
Kyoto 606-8502, Japan}

\author{\footnotesize MASAKIYO KITAZAWA\footnote{E-mail:
kitazawa@quark.phys.bnl.gov}}

\address{RIKEN BNL Research Center, Brookhaven Natianl Laboratory,
Upton, NY 11973,  USA}

\author{\footnotesize TEIJI KUNIHIRO\footnote{E-mail:
kunihiro@yukawa.kyoto-u.ac.jp}}

\address{Yukawa Institute for Theoretical Physics,
Kyoto 606-8502, Japan}

\author{\footnotesize YUKIO NEMOTO\footnote{E-mail:
nemoto@hken.phys.nagoya-u.ac.jp}}

\address{Department of Physics, Nagoya University
Nagoya 464-8602, Japan}

\maketitle

\begin{history}
\received{}
\revised{}
\end{history}

%%%%%abstract
\begin{abstract}
We calculate the spectral function of the massive quark
at finite temperature (T) using a Yukawa model and show that 
the peak in the negative energy region among the three-peaks 
found in a previous work for the massless quark is largely suppressed.
To explore the underlying mechanism of this behavior, 
we also investigate the pole structure of the retarded
Green function of the quark.
We will show the result only for the massless quark.
We find the residues of the poles corresponding 
the three-peaks are all comparable at $T\sim m_b$.
We also show that the multi-peak structure of the quark spectra
is well described in the pole approximation
which indicates that the quasi-paricle picture is valid
in this $T$ region.
\end{abstract}

%%%%%body of text
\section{Introduction}
The recent analyses of the experiment 
in the Relativistic Heavy Ion Collider (RHIC)
suggest
that quark-gluon plasma (QGP) near and above 
the critical temperature $T_c$ of QCD phase transition
is a rather strongly coupled system.
Such a picture is consistent with 
the lattice calculations\cite{Umeda,AH,Datta} and 
model calculations\cite{SZ,Brown,MR} 
which suggest the existence of hadronic excitations 
even above $T_c$.
In fact, the possible existence of hadronic excitations
above $T_c$ itself had been suggested earlier by
Hatsuda and Kunihiro\cite{HK}: 
They showed using an effective chiral model
that hadronic $\sigma$- and $\pi$-like excitations appear
as the soft mode of the chiral transition near but above $T_c$

Recently Kitazawa et al.\cite{KKN,KKN2} showed that
the quarks may have a collective nature and 
show also an anomalous behavior at $T$ near but
above $T_c$ in the chiral limit:
The quark coupled to those soft modes give rise
to a three-peak structure in the spectral function.
They also clarified with use of a Yukawa model
that the novel structure of the quark spectra is
owing to the level mixing between
quark (anti-quark) states and anti-quark hole (quark hole)
states in the thermally excited anti-quark (quark) distribution
via the ``resonant scattering'' of quarks with the soft modes.

In this report, we further examine their results
on the anomalous behavior of the quarks focusing 
on the pole structure of the collective quark excitations 
and the possible effects of the finite quark mass.
The relevance of our new findings of the collective nature
of the quarks to the RHIC experiments will be briefly discussed.

\section{Formulation}
We start from the following Lagrangian composed of 
a massive quark field $\psi$ and a massive scalar boson $\phi$:
\begin{equation}
\mathcal{L} = \bar{\psi}(i\Slash{\partial}-m_f-g\phi)\psi
+\frac{1}{2}\left(\partial_\mu\phi\partial^\mu\phi-m_b^2\phi\right),
\end{equation}
where $g$ is the coupling constant, $m_f$ the quark mass 
and $m_b$ the boson mass.

In this report, we focus on the case of a vanishing
external momuntum of the quark.

The quark self-energy
in the imaginary-time formalism 
at one-loop order is given by
\begin{equation}
\tilde{\Sigma}(i\omega_m)
=-g^2T\sum_{n}\int\frac{d^3\mathbf{k}}{(2\pi)^3}
\mathcal{G}_0(\mathbf{k},\omega_n)
\mathcal{D}(-\mathbf{k},i\omega_m-i\omega_n),
\end{equation}
where 
$\mathcal{G}_0(\mathbf{k},i\omega_n)
=[i\omega_n\gamma_0-(\mathbf{k}\cdot\mathbf{\gamma}+m_f)]^{-1}$
and 
$\mathcal{D}(\mathbf{k},i\nu_n)
=[(i\nu_n)^2-\mathbf{k}^2-m_b^2]^{-1}$ 
are Matsubara Green functions for the free quark and 
the scalar boson, respectively, with 
$\omega_n=(2n+1)\pi T$ and $\nu=2n\pi T$ being
the respective Matsubara frequencies.
One can get the retarded self-energy $\Sigma\reta(\omega)$
by carrying out the analytic continuation 
$i\omega_m \rightarrow \omega+i\eta$ after summing over
the Matsubara mode $n$.

The spectral functions of the quark and anti-quark
are expressed with the retarded Green function $G\reta(\omega)$ as 
$\rho_{\pm}(\omega)\equiv -(1/\pi) 
\tr[\im G\reta(\omega)\gamma_0\Lambda_\pm]
\equiv -(1/\pi)\im G\reta_\pm(\omega) $
where $\Lambda_\pm = (1\pm\gamma_0)/2$ are 
the number projection operators at vanishing momentum
and $G\reta_\pm(\omega)$
the quark and anti-quark components of the retarded Green functions.

We analytically continue the retarded Green function of 
the quark to whole complex energy plane and
search poles $z_+$ of the Green function by solving following equation
in the lower-half plane:
\begin{equation}
\left[G\reta_+(z_+)\right]^{-1}=0,
\label{eq:complex dispersion}
\end{equation}

The pole $z_+$ can be expressed as
\begin{equation}
z_+ = M - i\Gamma/2,
\end{equation}
where $M$ is the mass of the excitation and $\Gamma$ the width.

The residue at the pole is given by
\begin{equation}
Z=\lim_{z\rightarrow z_+}(z-z_+)G\reta_+(z),
\end{equation}
which represents the strength of the excitation.

\section{The Spectral function of massive quark}
In Fig.~\ref{fig:SpFnc}, we plot the $T$-dependence of
the spectral functions of the massless and massive quarks
\footnote{The spectral function for the massless quark in a Yukawa model
have been already given by Kitazawa et al.\cite{KKN2}.}.

\bgfig
\begin{tabular}{lr}
\begin{minipage}{0.5\hsize}
\psfig{file=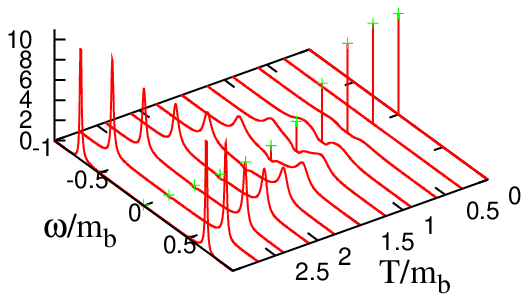,width=7cm}
\end{minipage}
\begin{minipage}{0.5\hsize}
\psfig{file=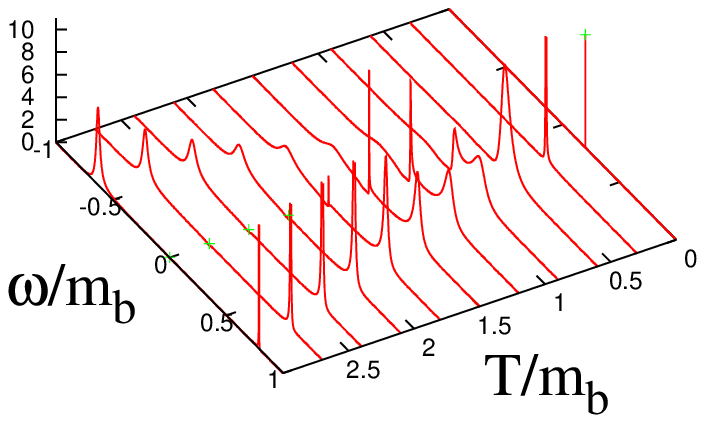,width=6.5cm}
\end{minipage}
\end{tabular}
\caption{The $T$-dependence of the spectral functions
of the massless quark (left panel) and the massive quark with
$m_f=0.2m_b$ (right panel). 
Each red line is the spectral function of the quark
with vanishing momentum at each $T$.
We approximate extremely narrow peaks by $\delta$-functions.
We express the $\delta$-functions by the vertical lines
with the heights which is ten-times of the residues at the peaks.
The Green points are the heads of the lines for $\delta$-functions.}
\label{fig:SpFnc}
\enfig

The spectral function for the massless quark has three-peak structure
at $T\sim m_b$.
We call the modes corresponding to the peaks in finite $\omega$
region the massive-mode and 
the mode corresponding to the peak at the origin
the massless-mode%\footnote{These masses are thermal mass.}.
At high $T$, the massive-modes in the positive and 
negative energy region connect to the normal quasi-quark mode
and the `plasmino' mode in the hard thermal loop (HTL) approximation,
respectively.
On the other hand, the massive-mode peak in the negative energy
region is suppressed in the case of the massive quark,
and accordingly the three-peak structure hardly remain.
This behavior is consistent with the interpretation 
of the multi-peak structure in the quark spectrum 
from the aspect of the level mixing in the previous
work\cite{KKN,KKN2} for the massless quark;
the quark level is shifted above in energy by quark mass, 
and hence the level mixing in the negative energy region
is suppressed.

We now give a suggestion on 
the quark spectrum near and above the critical temperature 
of the chiral phase transition.
We assign the ratio of parameters from a calculation in the NJL model\cite{HK},
because the model which we use have no phase transition.
The extracted ratio of the parameters are 
$T/m_b\sim 0.8$ and $m_f/m_b\sim 0.2$
at $T\sim 1.1T_c$. 
The spectral function qualitatively differ from 
both of zero temperature one and high temperature one.
The spectrum of the massive quark at $T\sim T_c$ may have a peak 
near the origin separately from the peak in the 
positive-energy region.

\section{The Pole structure of quark propagator}
We investigate the pole structure of the quark propagator
in order to elucidate the underlying mechanism of the
behavior of the spectral function shown in the preceding section.
Although we should show the result not only for the massless quark
but also the massive quark, 
the research in the later case is now in progress.
So we only show the result for the massless quark.
We focus on the poles which correspond to the peaks in
the spectral function of the quark, 
although it seems that there are some poles which do
not correspond to any peak.

A pole corresponding to the massless-mode stays
at the origin of the complex plane irrespective of $T$.
The poles of the massive-mode move with $T$ as 
shown in Fig~\ref{fig:glP}:

\bgfig
\centerline{\psfig{file=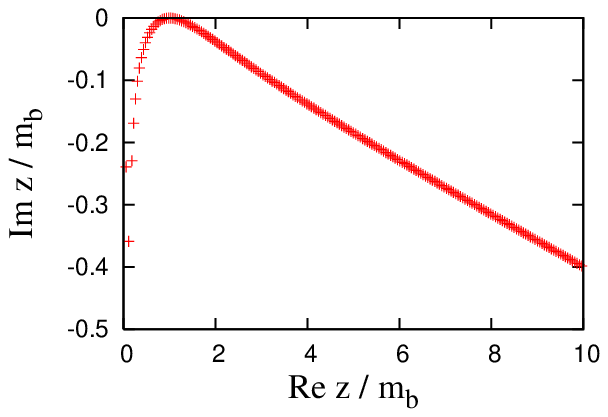,width=7cm}}
\caption{The $T$-dependence of the pole 
corresponding to the massive-mode in $\re z > 0$ region.
Because of the vanishing quark mass, 
the pole structures in $\re z>0$ region and $\re z<0$ region
are symmetric.
As $T$ is raised from $T=0.2m_b$ to $T=40m_b$,
the pole denoted by the red point moves from
far left to far right.
}
\label{fig:glP}
\enfig

At low $T$, the poles have large widths and 
these large widths are consistent with the fact that there is no clear peak
in the finite energy region of the spectral function as shown 
in the left panel of Fig.~\ref{fig:SpFnc}.
As $T$ is raised, $\Gamma$ decreses while $M$ increses
until some $T$ where $\Gamma$ vanishes.
After then the pole moves to large 
width region as $T$ is raised further.
We notice that this behavior is consitent width
the HTL approximation;
the poles in the HTL approximation may
have finite imaginary part in the order of $g^2T$.

We plot the $T$-dependence of the
residues  at the poles corresponding to 
the massless-mode and the massive-mode 
in Fig.~\ref{fig:residues}.
\bgfig
\centerline{\psfig{file=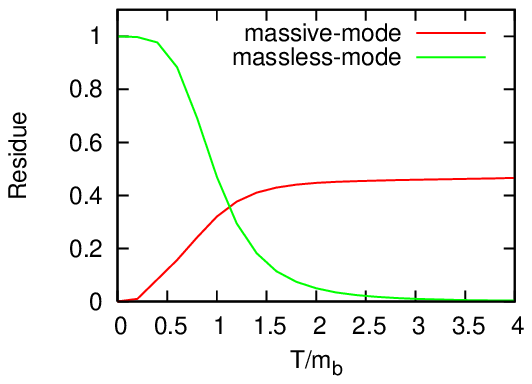,width=8cm}}
\caption{The $T$-dependence of residues.
The green line is for the massless-mode while 
the red line for the massive-mode.}
\label{fig:residues}
\enfig
The residues of the massless-mode and the massive-mode
are comparable with each other at $T\sim m_b$.

In the vicinity of the pole, the spectrum should be 
approximated by a Breit-Wigner type formula given by
\begin{equation}
\rho^\mathrm{BW}(\omega)=-\frac{1}{\pi}\im
\frac{Z}{\omega-z_+},
\label{eq:BW}
\end{equation}
where $z_+$ is the pole and $Z$ in the residue at $z_+$.

We plot the approximated spectral function in Fig.~\ref{fig:poleapp}.
\bgfig
\centerline{\psfig{file=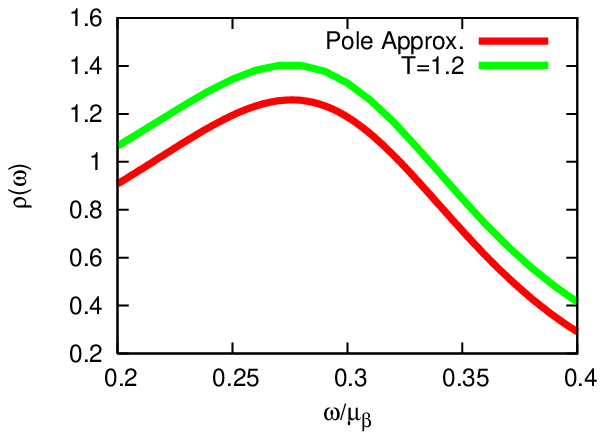,width=7cm}}
\caption{The pole approximation of the spectral function
at $T = 1.2m_b$. The red line is the 
approximated spectral function given by Eq.~\ref{eq:BW}.
while the green line the one without the approximation.}
\label{fig:poleapp}
\enfig
Comparing it with the spectral function without the approximation,
one can say that
the approximation by a Breit-Wigner type formula is works well,
which supports the validity of the quasi-particle picture of 
the quarks at $T\sim m_b$.

\section{Summary}
We have calculated the spectral function of the massive quark
using a Yukawa model and found that the peak in the negative
energy region is suppressed.
This behavior is consistent with the interpretation of the
multi-peak structure in the quark spectra given by
Kitazawa et al.\cite{KKN,KKN2}.
Employing the results given in the NJL model,
it is suggested that the spectral function
of the massive quark at $T\sim T_c$ may have a peak near the origin
separately from the peak in the positive-energy region.
We have also examined the pole structure of the retarded
Green function of the massless quark.
The result show that each of the all three peaks are
equally significant and the quasi-particle picture is 
quite valid at $T\sim m_b$.

M. K. is supported by a Special Postdoctoral Research Program of RIKEN.
T. K. is supported by a Grant-in-Aid for Scientific Research
by Monbu-Kagakusho of Japan (No 17540250).
Y. N. is supported by the 21st Century COE Program
at Nagoya University and
a Grand-in-Aid for Scientific Research by Monbu-Kagakusho of Japan
(No. 18740140).
This work is supported by the Grant-in-Aid
for the 21st Century COE ``Center for Diversity and Universality
in Physics'' of Kyoto University.

\end{document}